\documentclass[useAMS,usenatbib]{mn2e}
\usepackage{graphicx, subfig, amssymb, amsmath}

\def\kpnothanks{Visiting Astronomer, Kitt Peak National Observatory,
  National Optical Astronomy Observatory, which is operated by the
  Association of Universities for Research in Astronomy (AURA) under
  cooperative agreement with the National Science Foundation. }

\def\kms{km$\;$s$^{-1}$}
\def\Rproj{R_{\rm proj}}
\def\Msun{${\rm M}_{\odot}$}
\def\PSone{\emph{PanSTARRS1}}

\def\ds{\displaystyle}

\def\rin{r_{\rm in}}
\def\rout{r_{\rm out}}

\title[The Globular Cluster System of NGC~6822] {The Globular Cluster
  System of NGC~6822} \author[Veljanoski et al.]
{J.~Veljanoski$^{1,2}$, A. M. N.~Ferguson$^{1,}$\thanks{\kpnothanks},
  A. D.~Mackey$^3$, A. P.~Huxor$^4$, J. R.~Hurley$^5$,\ \newauthor
  E. J.~Bernard$^1$, P.~C\^{o}t\'{e}$^{6,\star}$, M. J.~Irwin$^7$,
  N. F.~Martin$^{8,9}$, W. S.~Burgett$^{10}$, \ \newauthor K. C.~Chambers$^{10}$,
  H.~Flewelling$^{10}$ R.~Kudritzki$^{10}$, and C.~Waters$^{10}$
  \\
  $^1$ Institute for Astronomy, University of Edinburgh, Royal Observatory, Blackford Hill, Edinburgh, EH9 3HJ, UK\\
  $^2$ Kapteyn Astronomical Institute, University of Groningen, P.O. Box 800, 9700 AV Groningen, The Netherlands \\
  $^3$ Research School of Astronomy $\&$ Astrophysics, Australian National University, Mt. Stromlo Observatory, Cotter Road,\\ Weston Creek, ACT 2611, Australia\\
  $^4$ Astronomisches Rechen-Institut, Zentrum f\"{u}r Astronomie der Universit\"{a}t Heidelberg, M\"{o}nchhofstra{\ss}e 12-14, 69120 Heidelberg, \\Germany.\\
  $^5$ Centre for Astrophysics $\&$ Supercomputing, Swinburne University of Technology, Hawthorn, VIC 3122, Australia\\
  $^6$ NRC Herzberg Institute of Astrophysics, 5071 West Saanich Road, Victoria, BC, V9E 2E7, Canada\\
  $^7$ Institute of Astronomy, University of Cambridge, Madingley, Road, Cambridge, CB3 0HA, UK \\
  $^8$ Observatoire astronomique de Strasbourg, Universit\'{e} de Strasbourg, CNRS, UMR 7550, 11 rue de l’Universit\'{e}, F-67000 Strasbourg, France \\
  $^9$ Max-Planck-Institut f\"{u}r Astronomie,  K\"{o}nigstuhl 17,  D-69117 Heidelberg, Germany\\
  $^{10}$ Institute for Astronomy, University of Hawaii at Manoa, Honolulu, HI 96822, USA}
\date{Draft version \today.}

\pagerange{\pageref{firstpage}--\pageref{lastpage}} \pubyear{2015}

\def\LaTeX{L\kern-.36em\raise.3ex\hbox{a}\kern-.15em
    T\kern-.1667em\lower.7ex\hbox{E}\kern-.125emX}

\begin{document}

\label{firstpage}

\maketitle

\begin{abstract}
  We present a comprehensive analysis of the globular cluster (GC)
  system of the Local Group dwarf irregular galaxy NGC~6822. Our study
  is based on homogeneous optical and near-IR photometry, as well as
  long-slit spectroscopic observations which are used to determine new
  radial velocities for 6 GCs, two of which had no previous
  spectroscopic information.  We construct optical-near IR
  colour-colour diagrams and through comparison to simple stellar
  population models infer that the GCs have old ages consistent with
  being 9~Gyr or older, while their metallicities are in the range
  between ${\rm -1.6 \lesssim [Fe/H] \lesssim -0.4}$.
  We conduct a kinematic analysis of the GC population and find
  tentative evidence for weak net rotation of the GC system, in the
  same sense as that exhibited by the underlying spheroid. The most
  likely amplitude of rotation is $\approx 10$ km$\;$s$^{-1}$,
  approximately half the magnitude of the observed velocity
  dispersion. Finally, we use the GCs to estimate the dynamical mass
  of NGC~6822 within $\sim 11$~kpc and we formally find it to be in
  the range between $(3-4) \times 10^{9}~M_{\odot}$. This implies an overall
  mass-to-light ratio in the range of $\sim 30 - 40$ and indicates that
  NGC~6822 is highly dark matter dominated. The mass and the corresponding
  mass-to-light ratio estimates are affected by various additional systematic
  effects due to limitations of the data and the model that are not necessary
  reflected in the formal uncertainties.
\end{abstract}


\section{Introduction}

According to the $\Lambda$CDM cosmological model, massive galaxies
build up in part through the amalgamation and merger of smaller
galaxies. The globular cluster (GC) systems of such massive galaxies,
which are often quite rich, are also thought to be assembled, at least
partially, from GCs donated by the smaller accreted galaxies
\citep[e.g.][]{Cote98,Cote00a}. If this is indeed the case, GCs can act as
valuable tracers of such events, while offering the potential to understand
the formation history of their host galaxy. In their influential study,
\citet{SZ78} uncovered important clues about the formation of the Milky Way
through its GC system, namely that accretion played a significant part in the
process.  Modern studies of the Galactic GC system arrive at the same conclusion
\citep[e.g.][]{Zinn93GC,Mackey04,Mackey05,MF09,Dotter10,Dotter11,Keller12}.
Indeed the growth of our Galaxy and its GC system is still ongoing.
Observations of the Sagittarius dwarf galaxy, which is currently being
accreted onto the Milky Way while donating at least 5 GCs to the
Galactic halo, provide the strongest evidence that this is the case
\citep[e.g.][]{Bellazzini03,Law10}. The work of our group has shown
that a similar situation is also observed in M31. The halo of this
galaxy features a number of prominent stellar streams with which many
GCs are associated, bearing witness to an ample accretion history
\citep[e.g.][]{Mackey10b,Mackey13,Mackey14,Huxor11,Veljanoski13a,Veljanoski14}.

The study of dwarf irregular galaxies (dIrr) is a particularly
important topic within the context of understanding galaxy formation.
Being by far the most common galaxy type at high redshift
\citep[e.g.][]{Stiavelli04,vanderWell11}, dIrr systems could have provided
a significant contribution toward the growth of massive galaxies at early
times. In addition, in order to use GC as probes of galaxy assembly,
it is crucial to understand the properties of GC systems in dwarf galaxies,
and how they relate to the host galaxy properties and to the properties of
GCs formed \emph{in situ} around a massive galaxy. In this context, dIrr
galaxies are especially helpful, since in the local universe they are often
found either in isolation or in groups without a massive dominating
companion, making them excellent systems for studying the pristine
properties of GC systems.

Lying in the southern hemisphere with Galactocentric coordinates of
$l=25.4^{\circ},b=-18.4^{\circ}$, NGC~6822 is one of the most
intriguing dwarf galaxies in the Local Group.  Discovered by
\citet{Barnard84}, this barred dIrr galaxy features a number of
peculiar properties that has made it the focus of much attention over
the years, even though its low Galactic latitude makes observations
quite challenging. It has an absolute magnitude ${M_V = -15.2}$ and a
half-light radius of $\sim 0.5$~kpc \citep{Hodge77,Hodge91}.
Located at a distance of $472$~kpc \citep{Gorski11}, it is the closest
dIrr galaxy save for the Magellanic Clouds. It does not appear to be
associated with either the Milky Way or M31, and it has no other
neighbouring companions. This galaxy contains a substantial spheroidal
component, that has been traced via its constituent red giant branch
stars out to a radius of at least $\sim 5$~kpc \citep{Battinelli06}.
\citet{Demers06kin} used carbon stars in the spheroid as kinematic
tracers, finding evidence for rotational support. In addition,
NGC~6822 contains large quantities of HI gas forming a disc, the
semi-major axis of which is positioned perpendicular to the stellar
spheroid \citep[e.g][]{deBlok00,deBlok06a,Weldrake03}. The HI gas is
also observed to be rotating, but at right angles to the rotation of
the spheroid, showing that there are at least two different kinematic
components present in NGC~6822. This is further supported by a recent
study on planetary nebulae and HII regions by \citet{Flores14},
showing that the planetary nebulae have kinematics similar to the
carbon stars, while the HII regions share the kinematics of the HI
disc.

\begin{figure}
\centering
\includegraphics[width=0.52\textwidth]{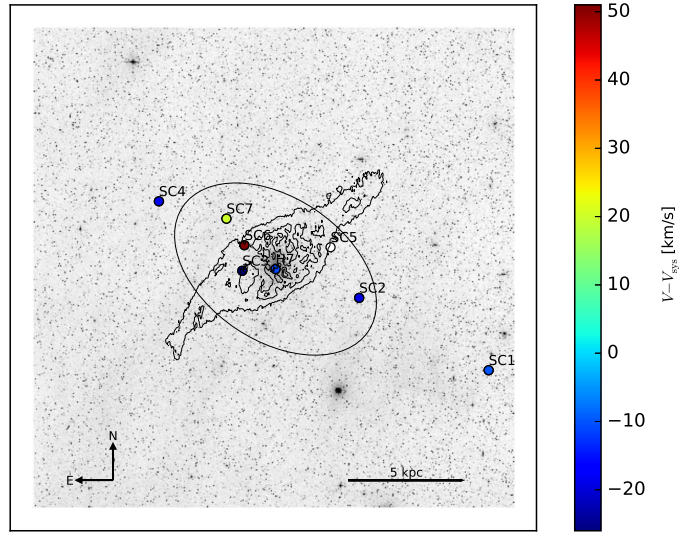}
\caption{Positions of the 8 known GCs overlaid on a \emph{Digital Sky
    Survey} image of NGC~6822 . The colours of the points correspond
  to their radial velocities, as measured in the present work (see
  Section \ref{s:kinematics}) and corrected for the systemic motion of
  NGC~6822.  The ellipse shows the extent of the stellar spheroid
  traced by \citet{Battinelli06}, with a semi major axis of 36~arcmin
  and an ellipticity of 0.36. Also shown are contours of the HI
  distribution map of \citet{deBlok00} ranging from 3.5 - 42 $\times 10^{20}$~cm$^{-2}$.}
\label{fig:spatialN6822}
\end{figure}

In recent years, dedicated searches have substantially increased the
number of GCs known around Local Group galaxies
\citep[e.g.][]{Huxor08, Veljanoski13b}, including NGC~6822.
Hubble~VII, originally discovered by \citet{Hubble25N6822}, was long
considered the only old GC in NGC~6822. However, this changed when
\citet{hwang11} discovered 4 new GCs in wide-field \emph{CFHT/MegaCam}
imagery covering an area of $3\times3$~deg around NGC~6822, .  Later,
\citet{huxor12} re-examined this data and by complementing it with
additional archival \emph{Subaru} and \emph{CFHT} imagery which
expanded the observed footprint, uncovered 3 additional GCs, bringing
the total number of members to 8. Figure~\ref{fig:spatialN6822} shows
the positions of the GCs superimposed on an image of NGC~6822, on
which are also marked the contours of the HI distribution, ranging between
3.5 - 42 $\times 10^{20}$~cm$^{-2}$
\citep{deBlok00}, and the known extent of the red giant branch stars in
the galaxy spheroid \citep{Battinelli06}. It is notable that the GC
system of NGC~6822 is quite spatially extended, with the most remote
cluster, SC1, having a projected radius of nearly 11~kpc.
Furthermore, the GCs are arranged in a rather linear configuration,
with an orientation aligned quite closely with the major axis of the
spheroid \citep{hwang11,huxor12}.  Five of the 8 GCs in NGC~6822 are
fairly extended in nature and thus resemble the extended clusters
found in the haloes of M31 \citep[e.g.][]{Huxor05,Huxor14} and M33
\citep{Stonkute08}. The coordinates, position angles, projected radii
($\Rproj$), GC type and Galactic $E(B-V)$ extinction values
\citep{Schlegel98} of the clusters are listed in
Table~\ref{tab:N6822coords}.

\begin{table*}
\centering
\footnotesize
\caption{Coordinates, position angles, projected radii, GC type (compact or extended),
  the half light radius (if known) and colour excess values due to Galactic
  interstellar reddening of the GCs in NGC~6822. The adopted centre of NGC~6822
  is RA = $19^h~44^m~57.7^s$, Dec = $-14^\circ~48'~12''$.}
\label{tab:N6822coords}
\begin{tabular}{lccccccccl}
  \hline
  \hline
  ID         & RA (J2000) & Dec (J2000) & $\Rproj$ & $\theta$ & GC type & $r_{\rm h}$  & $E(B-V)$ & Reference  \\
             & [h  m  s]  & [d  m  s]   & [kpc]    & [deg]    &         & [pc]         & [mag]    & \\
  \hline
  Hubble-VII & 19 44 55.8 & -14 48 56.2 & 0.1      & 227      & C       &  2.5$\pm$0.1 & 0.24     & \citet{Hubble25N6822} \\
  SC1        & 19 40 11.9 & -15 21 46.6 & 10.7     & 244      & E       & 14.0$\pm$0.2 & 0.16     & \citet{hwang11} \\
  SC2        & 19 43 04.5 & -14 58 21.4 & 4.1      & 250      & E       & 11.5$\pm$0.2 & 0.22     & \citet{hwang11} \\
  SC3        & 19 45 40.2 & -14 49 25.8 & 1.4      & 94       & E       &  7.5$\pm$0.5 & 0.19     & \citet{hwang11} \\
  SC4        & 19 47 30.4 & -14 26 49.3 & 6.0      & 60       & E       & 13.8$\pm$0.3 & 0.19     & \citet{hwang11} \\
  SC5        & 19 43 42.3 & -14 41 59.7 & 2.7      & 290      & E       & ...          & 0.22     & \citet{huxor12} \\
  SC6        & 19 45 37.0 & -14 41 10.8 & 1.6      & 52       & C       & ...          & 0.19     & \citet{huxor12} \\
  SC7        & 19 46 00.7 & -14 32 35.0 & 3.0      & 43       & C       & ...          & 0.21     & \citet{huxor12} \\
  \hline
\end{tabular}
\end{table*}

In this paper we present a comprehensive analysis of the GC system in
NGC~6822, based on uniform optical and near IR photometry and
long-slit spectroscopy.  The photometry is used to constrain the ages
and metallicities. Even though deeper optical data exist for some
clusters \citep[e.g.][]{hwang11,huxor12}, the uniformity of the data
presented in this study ensures no systematic offsets are present in
the age and metallicity estimates due to heterogeneous imaging. The
spectroscopy yields new radial velocity measurements for 6 of the GCs,
two of which had no previous velocity information. These data are also
used to redetermine the dynamical mass and mass-to-light ratio of this
galaxy.

\section{The Data}

\subsection{Optical data}

The optical imagery has been obtained as part of the \PSone\, 3$\pi$
Survey (Kaiser et al. 2010; K. Chambers et al., in preparation).  This
survey has targeted three quarters of the sky ($\delta > -30^{\circ}$)
in five optical bands, $g_{P1},r_{P1},i_{P1},z_{P1},y_{P1}$, with the
1.8~m \PSone\ telescope located on Haleakala, Hawaii
\citep{Tonry12photsys}.  The optical system of \PSone\, features a 1.4
Gigapixel imager \citep{Onaka08,Tonry09} with a field of view of 7
deg$^2$. The sky has been observed up to four times a year in each
band, with individual exposures being between 30 and 45s in length.
The data are automatically processed in real time with the {\sc Image
  Processing Pipeline} \citep[{\sc ipp},][]{Magnier06}.  After the
standard reduction processes such as flat-fielding, the pipeline
resamples all images to a uniform pixel size of of 0.25~arcsec and
aligns them to the equatorial axes onto ``skycells'' -- regular
patches on the sky of 6250 pixels across. The data are calibrated to
each other self-consistently using partially overlapping exposures
\citep{Schlafly12}, yielding a calibration precision better than 10
mmag as measured against SDSS.  For the work presented here, we use
data from the processing version 2 (PV2) and 2~$\times$~2~arcmin
cutouts of stacked images of the GCs provided by the postage stamp
server.

\subsection{Near-IR data}

\begin{figure*}
\centering
\includegraphics[scale = 0.35]{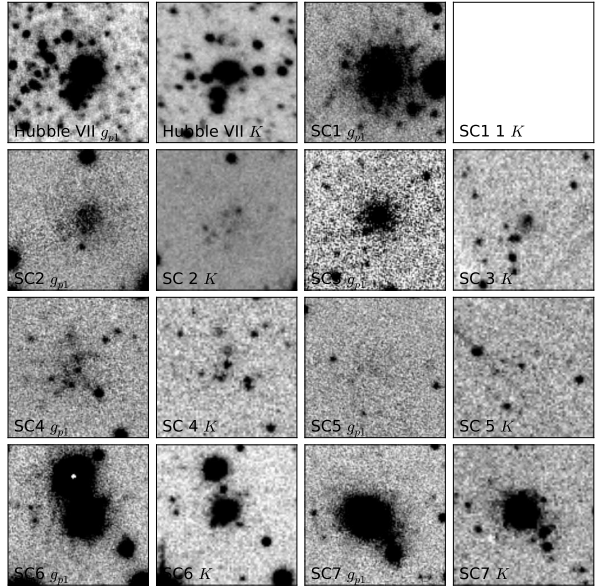}
\caption{$g_{P1}$ and $K$ band images of the GCs in NGC~6822. Each
  image is 60 $\times$ 60~arcsec wide. North is up and east is left.
  SC1 falls outside the near-IR survey and hence no $K$-band image is
  shown, while SC5 is only marginally detected in both passbands.}
\label{fig:N6822cutouts}
\end{figure*}

The near-IR imaging of NGC~6822 was taken as part of a project to
survey red stellar populations of Local Group galaxies. The data were
obtained in October 2008 using the \emph{WFCAM} instrument
\citep{Casali07} mounted on \emph{UKIRT}. \emph{WFCAM} has four
individual detectors, each with a pixel scale of 0.4~arcsec, arranged
such that four dithered pointings cover a square of 0.75 deg$^{2}$
when properly aligned. Our observations were conducted in three
near-IR bands $J$, $H$ and $K$. The data were processed with a
pipeline created by the Cambridge Astronomy Survey
Unit\footnote{http://casu.ast.cam.ac.uk}, which carries out the
standard dark current correction, flat-fielding, crosstalk removal,
systematic noise and sky removal. The pipeline, described in
\citet{Cioni08} and \citet{Hodgkin09}, also performs full astrometric
and photometric calibration, based on the \emph{2MASS} point source
catalogue. The reduced images are stacked and microstepped to produce
individual science frames. The finalised frames are not resampled, and
thus they retain the original pixel scale of the detector. The nightly
zero-point variation is $< 1$\%.  Figure~\ref{fig:N6822cutouts} shows
60 $\times$ 60~arcsec wide cutouts of each GC hosted by NGC~6822 in
the \PSone\, $g_{P1}$ and \emph{UKIRT/WFCAM} $K$ bands. Note that SC1
falls outside the near-IR coverage and thus no thumbnail is shown,
while SC5 is only marginally detected in both optical and near-IR data
sets.

\subsection{Spectroscopic data}

Moderate resolution spectra for 6 GCs were acquired in the summer of
2009 using the \emph{ISIS} instrument mounted on the 4.2-m William
Hershel Telescope (\emph{WHT}), the \emph{RC}-spectrograph installed
on the 4-m Kitt-Peak National Observatory (KPNO) telescope, and the
\emph{DEIMOS} spectrograph used by the 10-m W. M. Keck telescope.
Several exposures of each GC were obtained, with individual
integration times determined according to the brightness of each
target. Our observing log is shown in Table~\ref{tab:specObsN6822}.

\begin{table*}
\centering
\footnotesize
\caption{Log of spectroscopic observations for GCs around NGC~6822}
\label{tab:specObsN6822}
\begin{tabular}{lccccl}
  \hline
  \hline
  ID                      & Date of obs.          & Number of             & Integration time      & Telescope    & Principal    \\
  &                       & exposures             & per exposure [s]      &               & investigator \\
  \hline
  Hubble VII              & 14/08/2009            & 3                     & 900                   & \emph{KPNO}   & A. Ferguson  \\
  SC1                     & 16/08/2009            & 3                     & 1800                  & \emph{KPNO}   & A. Ferguson  \\
  SC3                     & 17/08/2009            & 3                     & 1800                  & \emph{KPNO}   & A. Ferguson  \\
  SC6                     & 15/08/2009            & 3                     & 900                   & \emph{KPNO}   & A. Ferguson  \\
  \vspace{1.5mm}
  SC7                     & 15/08/2009            & 3                     & 600                   & \emph{KPNO}   & A. Ferguson  \\
  SC2                     & 16-17/08/2009         & 3                     & 1800                  & \emph{WHT}    & A. Huxor     \\
  \vspace{1.5mm}
  SC7                     & 16/08/2009            & 2                     & 600                   & \emph{WHT}    & A. Huxor     \\
  SC1                     & 26/06/2009            & 3                     & 300                   & Keck          & J. Hurley    \\
  SC2                     & 26/06/2009            & 3                     & 300                   & Keck          & J. Hurley    \\
  SC6                     & 26/06/2009            & 3                     & 300                   & Keck          & J. Hurley    \\
  SC7                     & 26/06/2009            & 3                     & 420                   & Keck          & J. Hurley    \\
  \hline
\end{tabular}
\end{table*}

\subsubsection{\emph{WHT} and \emph{KPNO} data}

The \emph{ISIS} instrument comprises two individual detectors,
attached to separate `arms' of the spectrograph, that independently
sample different wavelength ranges. The blue arm covers a wavelength
range between $\sim 3500 - 5100$~\AA\ with a dispersion of
0.4~\AA~pixel$^{-1}$, selected via the usage of the EE12 camera along
with the R600B grating. The resolving power is R~$\sim 1500$.
Conversely, the red arm uses the RED-PLUS detector together with the
R600R grating to select the wavelength range between $\sim7400 -
9200$~\AA\ with a dispersion of 0.49~\AA~pixel$^{-1}$. The resolving
power in this case is R~$\sim 2700$. A similar set up was used for the
\emph{KPNO} observations. The T2KB detector, in conjunction with the
KPC007 grating were employed in order to select the wavelength range
between $\sim 3500-6500$~\AA\ with a dispersion of
0.139~\AA~pixel$^{-1}$ and a resolving power of R~$\sim 1300$. The
spectra obtained with both telescopes were not binned in either the
spatial or the wavelength directions. The typical S/N of the spectra
is $\sim 4-20$ per \AA.

The reduction of the spectroscopic data was carried out in {\sc
  iraf}\footnote{{\sc iraf} is distributed by the National Optical
  Astronomy Observatories, which are operated by the Association of
  Universities for Research in Astronomy, Inc., under cooperative
  agreement with the National Science Foundation.}. The basic
reduction steps -- bias and overscan subtraction, flat-fielding,
illumination correction -- were done with appropriate tasks in the
{\sc ccd} package. Having completed the initial reduction steps on the
two-dimensional frames, the \emph{apall} task in the {\sc kpnoslit}
package was used to extract one-dimensional spectra from them. We set
the extraction aperture to be between 2-4~arcsec. With the same task,
we interactively selected appropriate background sky regions, which we
fit with a 2nd order Chebyshev polynomial and subtracted. The spectra
were then traced using a 3rd order cubic spline function, and
extracted using the optimal variance weighting option in \emph{apall}.

We established wavelength calibration of the one-dimensional frames
via Cu-Ne-Ar and He-Ne-AR lamps for \emph{WHT} and \emph{KPNO} spectra
respectively. Comparison `arcs' were acquired before and after each
cluster exposure. We extracted the arc spectra with an identical
strategy as the target GCs they were used to calibrate. Using the
\emph{identify} task, we pinpointed $\sim 90\ ISIS$ blue, $\sim 25\
ISIS$ red and $\sim 50\ RC$ lines in the arc spectra and fitted a
dispersion solution with a 3rd order cubic spline function. The
root-mean-square residuals of the fits are $0.05\pm0.01$~\AA,
$0.02\pm0.01$~\AA\ and $0.08\pm0.01$~\AA\ for the data obtained with
the \emph{ISIS} blue, \emph{ISIS} red and \emph{RC} instruments
respectively. Since two arc spectra were observed for each target
exposure, the resulting wavelength solutions were averaged before
being applied to the appropriate GC via the \emph{dispcor} task. To
assess the reliability of the wavelength calibration, we measured the
positions of sky emission lines in the sky spectra, which are a
by-product of the \emph{apall} extraction.  We found that the
wavelength calibration is accurate to 0.08~\AA\ for the \emph{WHT} and
0.1~\AA\ for the \emph{KPNO} data.

\begin{figure*}
\centering
\includegraphics[scale = 0.75]{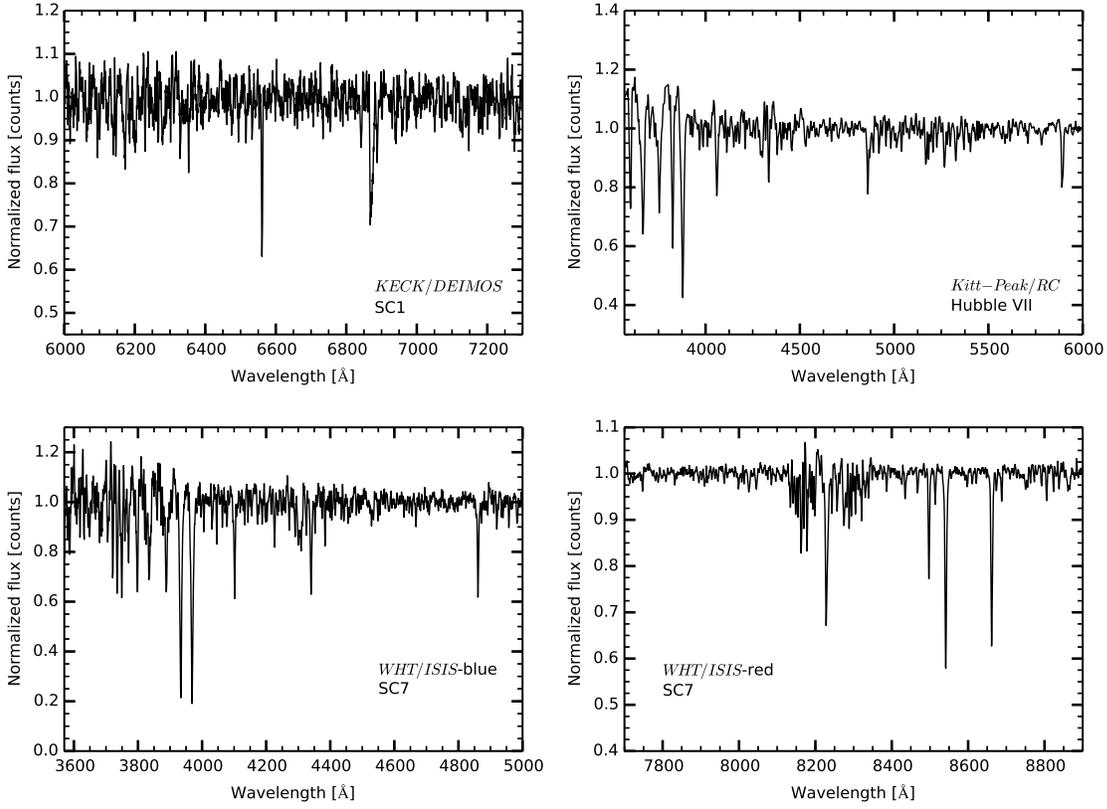}
\caption[Examples spectra of the NGC~6822 GCs]{Examples of fully reduced,
continuum-normalized spectra of GCs in NGC~6822.}
\label{fig:specExamplesN6822}
\end{figure*}

Given that all but one cluster were observed as a series of separate
exposures with each instrument, these were stacked using an inverse
variance weighting technique \citep[e.g.][]{Veljanoski14}.  Finally,
we continuum subtracted the spectra for the purpose of measuring
radial velocities.  Examples of the fully reduced spectra obtained
with each instrumental set up are shown on
Figure~\ref{fig:specExamplesN6822}. The displayed spectra are
continuum normalized rather than continuum subtracted in order to
preserve the relative strengths of the absorption lines for better
visualisation.

\subsubsection{Keck data}

We used the Deep Extragalactic Imaging Multi-Object Spectrograph
(\emph{DEIMOS}) in longslit mode to observe four NGC~6822 GCs (Swinburne
Keck program 2009A\_W006D, PI Hurley). The longslit mode of \emph{DEIMOS}
is achieved by inserting a standard single slit mask into the optical path
of the instrument. For all observations with this spectrograph we used a
mask featuring a slit 0.8~arcsec wide. The \emph{DEIMOS} camera comprises
of 8 CCD chips arranged in a $4 \times 2$ mosaic pattern, with the long
axis of this array oriented in the spatial direction and the short axis
in the dispersion direction. This setup essentially produces two spectra
simultaneously for a given compact target -- one on a `blue' CCD and
one on a `red' CCD.  We selected a central wavelength of 6040~\AA\ via
the 1200~l~mm$^{-1}$ grating, resulting in a dispersion of
0.33~\AA~pixel$^{-1}$. The resolving power of the spectra observed
with the red CCD is R~$\sim4700$.

To reduce observations of a given GC we first identified the relevant
CCD chips covered by the blue and red spectra.  We isolated these and
reduced them separately; the remaining 6 CCD chips were discarded, as
they contained no signal coming from our target.  Following this, the
standard reduction steps such as bias and overscan subtraction, and
flat-fielding, were undertaken with appropriate routines in the {\sc
  ccd} package in {\sc iraf}.  Inspecting the reduced 2-dimensional
frames, significant geometrical distortions in the dispersion of the
target objects are clearly evident, as well as distortions in the sky
lines. To rectify these we simply wavelength calibrated the
2-dimensional frames prior to extracting a 1-dimensional spectrum.

The wavelength solution is based on a Ne-Ar-Kr-Xe lamp, taken during
the baseline calibrations of the instrument. For the frames containing
the redder part of the spectrum, we identified 23 lines in the arc
spectrum via the \emph{identify} and \emph{reidentify} tasks. The
2-dimensional wavelength solution was then determined with the
\emph{fitcoords} task located in the {\sc twodspec} package. The
root-mean-square residuals of the fit were $0.02 \pm 0.01$~\AA. We
next applied the wavelength solution to the appropriate frames via the
\emph{transform} routine found in the same package. Inspecting the
wavelength calibrated frames, one sees that the distortions in the
spatial axis have disappeared, and the skylines are perfectly straight
in a direction perpendicular to the dispersion axis. This is a good
indicator that the wavelength calibration is reliable. For clarity,
the wavelength range of our spectra observed with the red chip spanned
$\sim6000 - 7250$~\AA.

Unfortunately, we were unable to wavelength calibrate the frames
containing the blue part of the spectra. This is because the arc
spectrum observed at these wavelengths contained only three reliable
lines, all of which sit extremely close to the edge of the frame. This
is a well known calibration issue for data taken in this specific
wavelength range (albeit one that is nowadays possible to cicumvent).
Even though this problem forced us to proceed further using only the
`red' frames, it did not hinder any of the velocity estimation and
kinematic analysis which we present in the later sections of this
paper.

From the appropriate 2-dimensional frames we extracted 1-dimensional
spectra and stacked these in the same manner as for the 4-m data, with
the only difference being the width of the extraction aperture, which
for the Keck data we set to 1.1~arcsec. As before, we used the sky
spectra associated with the extracted 1-dimensional spectra to assess
the accuracy of the wavelength calibration. We find that the
wavelength calibration is typically accurate to 0.06~\AA. An example
of a fully reduced, wavelength calibrated Keck spectrum is shown in
Figure~\ref{fig:specExamplesN6822}.

\section{Photometry}

We determined the total magnitudes of each GC in NGC~6822 via aperture
photometry in all 8 bands, $g_{P1}$, $r_{P1}$, $i_{P1}$, $z_{P1}$,
$y_{P1}$, $J$, $H$, and $K$. The photometric measurements were
conducted using the \emph{phot} task in {\sc iraf}. Accurate centres
for the classical, i.e. compact, GCs were established with the
centroid algorithm within \emph{phot}. Since the astrometric
calibration is identical for the entire \PSone\ data set we performed
the centring in the $r_{P1}$ band imaging, since the clusters are best
seen there. For the diffuse, extended clusters, we determined the
centre via visual inspection. Offsets from a nearby bright star were
used to accurately redetermine the centre of each cluster through the
rest of the data.

We used circular apertures to sum the total flux coming from each
cluster. The aperture size for a given cluster was chosen such that
the aperture enclosed the entire extent of the target; these sizes are
listed in Table~\ref{tab:N6822phot}. The background sky was determined
between two concentric circular annuli surrounding the photometric
aperture. We carefully placed these in a way so as to exclude unwanted
contaminants such as extended background galaxies or foreground Milky
Way stars. The reported magnitude uncertainties, which also include
the variation in the sky, are formally determined by the \emph{phot}
task.

All our photometric measurements were zero-point calibrated and
corrected for atmospheric extinction, and then subsequently corrected
for foreground reddening using $E(B-V)$ values from
\citet{Schlegel98}.  We calculated the extinction coefficients for the
\PSone\ bands using the relations presented in \citet{Tonry12photsys}:

\begin{align}
  Ag_{P1}/E(B{-}V) &= 3.613 - 0.0972 (g_{P1}{-}i_{P1}) + 0.0100 (g_{P1}{-}i_{P1})^2\\
  Ar_{P1}/E(B{-}V) &= 2.585 - 0.0315 (g_{P1}{-}i_{P1})                   \\
  Ai_{P1}/E(B{-}V) &= 1.908 - 0.0152 (g_{P1}{-}i_{P1})                   \\
  Az_{P1}/E(B{-}V) &= 1.499 - 0.0023 (g_{P1}{-}i_{P1})                   \\
  Ay_{P1}/E(B{-}V) &= 1.251 - 0.0027 (g_{P1}{-}i_{P1})
\end{align}
\citet{Schlafly11} recalibrated the $E(B-V)$ from \citet{Schlegel98}
and recommend that the \PSone\ reddening coefficients be further
multiplied by a factor of 0.88, which we included in the magnitude
derivation. Finally, Table~\ref{tab:N6822phot} lists the
extinction-corrected total magnitudes of all GCs in NGC~6822.

\begin{table*}
\centering
\scriptsize
\caption{Optical and near-IR photometry of all GCs hosted by NGC~6822.
The superscript (1) denotes that all magnitudes for this cluster are lower limits (i.e., upper limits to the luminosity).}
\label{tab:N6822phot}
\begin{tabular}{@{}lccccccccc@{}}
\hline
\hline
ID  &          Aperture  &   $g_0$  &              $r_0$  &              $i_0$  &              $z_0$  &              $y_0$  &              $J_0$  &              $H_0$  &              $K_0$  \\
  &            [arcsec]  &   [mag]  &              [mag]  &              [mag]  &              [mag]  &              [mag]  &              [mag]  &              [mag]  &              [mag]  \\
\hline
Hubble VII &   3.7 &         15.46 $\pm$ 0.02 &    14.99 $\pm$ 0.02 &    14.72 $\pm$ 0.02 &    14.58 $\pm$ 0.03 &    14.47 $\pm$ 0.04 &    13.44 $\pm$ 0.01 &    12.98 $\pm$ 0.02 &    12.86 $\pm$ 0.01 \\
SC1 &          7.0 &         16.62 $\pm$ 0.02 &    16.19 $\pm$ 0.02 &    15.97 $\pm$ 0.02 &    15.83 $\pm$ 0.03 &    15.72 $\pm$ 0.04 &    ...              &    ...              &    ...              \\
SC2 &          8.0 &         17.64 $\pm$ 0.02 &    16.98 $\pm$ 0.02 &    16.80 $\pm$ 0.02 &    16.63 $\pm$ 0.03 &    16.57 $\pm$ 0.05 &    15.53 $\pm$ 0.03 &    15.11 $\pm$ 0.03 &    14.86 $\pm$ 0.03 \\
SC3 &          3.0 &         18.79 $\pm$ 0.02 &    18.35 $\pm$ 0.02 &    18.08 $\pm$ 0.02 &    17.98 $\pm$ 0.03 &    17.77 $\pm$ 0.04 &    16.93 $\pm$ 0.04 &    16.44 $\pm$ 0.04 &    16.32 $\pm$ 0.05 \\
SC4 &          8.0 &         17.94 $\pm$ 0.02 &    17.37 $\pm$ 0.02 &    17.00 $\pm$ 0.02 &    16.84 $\pm$ 0.03 &    16.78 $\pm$ 0.02 &    15.74 $\pm$ 0.04 &    15.31 $\pm$ 0.04 &    15.11 $\pm$ 0.04 \\
SC5$^1$ &      7.5 &         $ \gtrsim 19.3 $ &    $ \gtrsim 18.6$  &    $ \gtrsim 18.2 $ &    $ \gtrsim 18.2 $ &    $ \gtrsim 17.8 $ &    $ \gtrsim 18.0 $ &    $ \gtrsim 16.6 $ &    $ \gtrsim 16.4 $ \\
SC6 &          4.7 &         15.76 $\pm$ 0.02 &    15.26 $\pm$ 0.02 &    15.03 $\pm$ 0.01 &    14.91 $\pm$ 0.03 &    14.83 $\pm$ 0.02 &    13.89 $\pm$ 0.01 &    13.42 $\pm$ 0.02 &    13.30 $\pm$ 0.01 \\
SC7 &          5.0 &         15.24 $\pm$ 0.02 &    14.60 $\pm$ 0.02 &    14.29 $\pm$ 0.01 &    14.12 $\pm$ 0.03 &    13.99 $\pm$ 0.02 &    12.93 $\pm$ 0.01 &    12.39 $\pm$ 0.02 &    12.30 $\pm$ 0.01 \\
\hline
\end{tabular}
\end{table*}

\begin{table*}
\centering
\scriptsize
\caption{Photometry of the NGC~6822 GCs converted to the Johnson/Cousins system using
  Equation~\ref{eq:p1-to-V} and \ref{eq:p1-to-I} taken from \citet{Tonry12photsys}.
  For convenience, the $(V-I)_0$ colours are also listed. The superscript is as in
  Table~\ref{tab:N6822phot}.}
\label{tab:N6822photVI}
\begin{tabular}{@{}lcccr@{}}
\hline
\hline
ID  &                   $V_0$  &                       $I_0$              & $(V-I)_0$              & $M_{V,0}$               \\
    &                   [mag]  &                       [mag]              & [mag]                  & [mag]                   \\
\hline
Hubble VII &            15.05  $\pm$ 0.02  &           14.18  $\pm$ 0.03  &  $0.87 \pm 0.04$       & $-8.3$                       \\
SC1 &                   16.30  $\pm$ 0.02  &           15.48  $\pm$ 0.03  &  $0.82 \pm 0.04$       & $-7.1$                       \\
SC2 &                   17.15  $\pm$ 0.03  &           16.25  $\pm$ 0.03  &  $0.90 \pm 0.04$       & $-6.2$                       \\
SC3 &                   18.44  $\pm$ 0.04  &           17.57  $\pm$ 0.03  &  $0.87 \pm 0.05$       & $-4.9$                       \\
SC4 &                   17.52  $\pm$ 0.03  &           16.47  $\pm$ 0.02  &  $1.05 \pm 0.04$       & $-5.9$                       \\
SC5$^1$ &               $ \gtrsim 18.8 $   &           $ \gtrsim 17.6 $   &  $ \sim 0.76   $       & $\gtrsim -4.6$               \\
SC6 &                   15.38  $\pm$ 0.02  &           14.51  $\pm$ 0.02  &  $0.87 \pm 0.03$       & $-8.0$                       \\
SC7 &                   14.77  $\pm$ 0.02  &           13.75  $\pm$ 0.02  &  $1.02 \pm 0.03$       & $-8.6$                       \\
\hline
\end{tabular}
\end{table*}

It is important to note that SC5 is at best marginally detected in the
imaging, and thus the measured values represent lower limits to its
magnitudes in all bands (i.e., upper limits to its luminosity).  The
extended cluster SC1 is the most remote object in this system and is
located outside the coverage of the near-IR data.

To allow for comparison with other studies, we convert the photometry
measurements from the \PSone\ system to the classical Johnson/Cousins
system using equations derived in \citet{Tonry12photsys}:

\begin{subequations}
\label{eq:p1-to-VI}
\begin{align}
{V} - r_{P1} = 0.005 + 0.462(g-r)_{P1} + 0.013(g-r)^2_{P1} \label{eq:p1-to-V} \\
{I} - i_{P1} = -0.366 - 0.136(g-r)_{P1} - 0.018(g-r)^2_{P1} \label{eq:p1-to-I}
\end{align}
\end{subequations}
The conversion to the Johnson $V$ and Cousins $I$ bands adds an
additional 0.012~mag and 0.017~mag to the uncertainty, respectively.
The results of the conversions are listed in
Table~\ref{tab:N6822photVI}. This table also lists the absolute
magnitudes ${M_{V_0}}$, calculated assuming each cluster has the same
distance of 472~kpc as the centre of NGC~6822. In
Table~\ref{tab:photcomp} we compare our photometric measurements to
those in previously published studies. In their discovery paper,
\citet{huxor12} presented optical photometric for SC6 and SC7,
measured on \emph{CFHT/MegaCam} archival images. There is an excellent
agreement between the photometry presented in this work to that in
\citet{huxor12}, with the measurements being less than one standard
deviation apart. Comparing the optical $(V-I)_0$ colour from our study
to those published in \citet{hwang11}, one finds mixed results. While
there is an excellent agreement between the colour measurements of our
present study and that of \citet{hwang11} for SC1 and SC2, there is a
poor agreement regarding the colour values of SC4, and even a strong
disagreement in the case of SC3. In addition, the colours presented by
\citet{hwang11} are always redder than those in this study.

\begin{table}
\centering
\scriptsize
\caption{$(V-I)_0$ colour comparison between the present study to those of \citet{huxor12}
  and \citet{hwang11}. }
\label{tab:photcomp}
\begin{tabular}{llll}
\hline
\hline
ID              & $(V-I)_0$              & $(V-I)_0$             & $(V-I)_0$               \\
                & This work              & \citet{huxor12}       & \citet{hwang11}         \\
\hline
Hubble-VII      & $0.87 \pm 0.04$        & $...          $       & ...                     \\
SC1             & $0.82 \pm 0.04$        & $...          $       & $0.85 \pm 0.03$         \\
SC2             & $0.90 \pm 0.04$        & $...          $       & $0.94 \pm 0.03$         \\
SC3             & $0.87 \pm 0.05$        & $...          $       & $1.31 \pm 0.03$         \\
SC4             & $1.05 \pm 0.04$        & $...          $       & $1.12 \pm 0.03$         \\
SC5             & $ \sim 0.76   $        & $...          $       & $...          $         \\
SC6             & $0.87 \pm 0.03$        & $0.84 \pm 0.03$       & $...          $         \\
SC7             & $1.02 \pm 0.03$        & $1.05 \pm 0.03$       & $...          $         \\
\hline
\end{tabular}
\end{table}

\section{Ages and Metallicities}

The best way of determining metallicities and ages of GCs is through
the analysis of deep colour-magnitude diagrams of their resolved
stellar populations. Second to that, high quality spectroscopic data
can also be used to derive the metal content and age of GCs. However,
not all of our spectra have sufficiently high S/N to reliably
determine these quantities.  Broadband colour-colour diagrams using a
combination of optical-optical and optical-near-IR colours are another
way to estimate ages and metallicities of GCs, and require only the
existence of homogeneous well-calibrated photometry \citep[e.g.][]{Puzia02,
Hempel03,Veljanoski13b,Munoz14}. The mechanism behind this is that both
$(g-i)_0$ and $(g-K)_0$ have similar sensitivity to the stars located near
the main sequence turn off point, which are most sensitive to age, but
$(g-K)_0$ is considerably more affected by the temperature of the red giant
branch which is more dependant on metallicity than age.  Hence, comparing
where GCs on such a plot sit relative to simple stellar population model
tracks allows a crude estimate of age and metal content of each cluster.

\begin{figure*}
\centering
\includegraphics[width=0.45\textwidth]{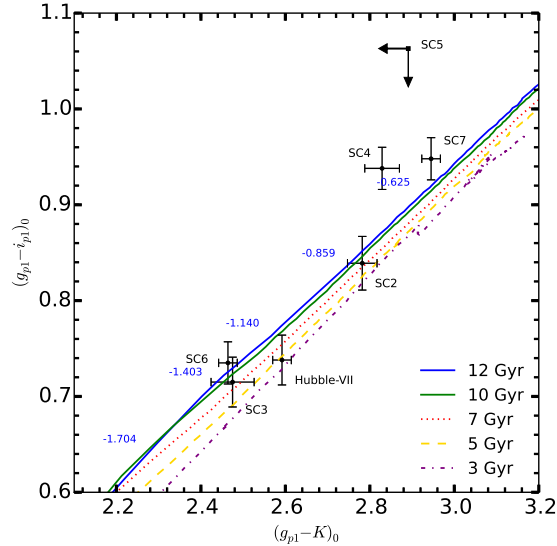} \\
\caption{$(g_{P1}-i_{P1})_0$ vs $(g_{P1}-K)_0$ colour-colour diagram
  of the GCs in NGC~6822, overlaid on top of the simple stellar
  population isochrones derived from the PARSEC1.1 models
  \citep{Bressan12}. Also shown are [Fe/H] values along the 12 Gyr
  isochrone. SC5 is represented by a square since this cluster is a
  marginal detection in the imaging. The clusters with reliable
  photometry are all found to have old ages.}
\label{fig:SSPN6822}
\end{figure*}

In Figure~\ref{fig:SSPN6822} we show a $(g_{P1}-i_{P1})_0$ vs
$(g_{P1}-K)_0$ colour-colour diagram constructed with the photometry
presented in Table~\ref{tab:N6822phot}. Integrated magnitudes, in
appropriate filters, of simple stellar populations having several
discrete ages are also overlaid on the same plot. They are derived
from the PARSEC1.1 models \citep{Bressan12}, and span the metallicity
range of ${\rm -2.2 < [Fe/H] < +0.6~dex}$. Formally, the diagram shows
that the clusters have old ages spanning the range between $\sim 5$ and
12~Gyrs, but they are all consistent with being 9~Gyrs or older. This is
in agreement with the spectroscopic study by \citet{Hwang14}.
The only exception is the cluster SC5, which being only marginally
detected at best in both the optical and near-IR imaging, does not lie
near the model tracks and thus its age and metallicity can not be
reliably constrained. For completeness, it is shown as a filled square
in Figure~\ref{fig:SSPN6822}.

We estimate the metallicities of the clusters via two different
methods. In both methods, we ignore SC5 since this cluster does not
have reliable photometry. In the first method, the metallicity of each
cluster is determined directly from Figure~\ref{fig:SSPN6822}. To do this,
we projected the position of each GC onto a grid of 12~Gyr isochrones in
color-color space and determined which track lay closest to the GC. The 12~Gyr
isochrone set is well sampled in metallicity and thus no interpolation is needed
for extracting the correct metallicities. The upper and lower uncertainties are
determined in the exactly the same way, considering the bluest and reddest
possible combination of colours of each clusters respectively, as allowed
by the colour uncertainties. The metallicities of the clusters determined
in this way are presented in the first column in Table~\ref{tab:cal6822}.

A different method of estimating GC metallicities is through empirical
colour-metallicity relations. Here we use the relation derived by
\citet{KP02}, calibrated using 129 Milky Way and M31 GCs that have
$E(B-V)<0.27$, to set additional constraints on the metallicities of
the NGC~6822 GCs. The relation is valid over a metallicity range of
$-2.3 <$ {\rm [Fe/H]} $< -0.2$~dex, and it has the form:
\begin{equation}
{\rm [Fe/H]} = (V-K)_0\times1.82\pm0.11 - 5.52\pm0.26
\label{eq:KP02}
\end{equation}
with an rms of 0.29 dex. Note that such a linear relation between colour and
metallicity is a crude assumption \citep{Cantiello07}.

The metallicities obtained via Equation~\ref{eq:KP02} are also shown in
Table~\ref{tab:cal6822}.  It is apparent that there is a modest
discrepancy between the estimated metallicities via the two different
methods. Although all measurements, except that for SC2, formally
agree with each other, a clear and nearly constant offset of $\sim
0.37$~dex is present. This mismatch is most likely due to systematics
in the stellar evolutionary models. Given how rapidly these models
progress, at the moment we find the metallicities estimated via the
empirical relation more reliable.

Table~\ref{tab:cal6822} also lists the [Fe/H] estimates from the
spectroscopic study of \citet{Hwang14}, as well as spectroscopic
metallicities for Hubble VII from earlier studies
\citep{Cohen98,Chandar00}.  In general, the photometric metallicities
derived here are considerably higher than those presented by
\citet{Hwang14}, with only SC3 being in formal agreement.  However, no
uniform offset is present between these two sets of
measurements. Furthermore, while our photometric [Fe/H] value for
Hubble VII is also higher than those derived in \citet{Cohen98} and
\citet{Chandar00}, the measurements agree within the reported
uncertainties.

\begin{table*}
\centering
\caption{[Fe/H] values for the GCs with reliable photometry in NGC~6822. Columns refer to:
  (a) values obtained directly from Figure \ref{fig:SSPN6822} by determining the projection of each
  cluster onto the 12~Gyr isochrone;
  (b) values obtained via Equation \ref{eq:KP02}, an empirical colour-metallicity relation
  calibrated on the M31 GC system \citep{KP02}; spectroscopic metallicity from (c)
  \citet{Hwang14}; (d) \citet{Cohen98}; (e) \citet{Chandar00}.}
\label{tab:cal6822}
\begin{tabular}{llllll}
\hline
\hline
ID         & [Fe/H](a)               & [Fe/H](b)      & [Fe/H](c)         & [Fe/H](d)         & [Fe/H](e)     \\
           & [dex]                   & [dex]          & [dex]             & [dex]             & [dex]         \\
\hline
Hubble-VII & $-1.14_{-0.09}^{+0.04}$ & $-1.5 \pm 0.4$ & $-2.34  \pm 0.03$ & $-1.95 \pm 0.15 $ & $ -2.0 \pm 0.25$          \\
SC1        & ...                     & ...            & $-2.00  \pm 0.04$ & ...               & ...                       \\
SC2        & $-0.84_{-0.06}^{+0.04}$ & $-1.3 \pm 0.4$ & $-2.53  \pm 0.06$ & ...               & ...                       \\
SC3        & $-1.34_{-0.15}^{+0.06}$ & $-1.7 \pm 0.4$ & $-1.52  \pm 0.06$ & ...               & ...                       \\
SC4        & $-0.75_{-0.05}^{+0.06}$ & $-1.1 \pm 0.4$ & $-2.53  \pm 0.08$ & ...               & ...                       \\
SC5        & ...                     & ...            & $ ...           $ & ...               & ...                       \\
SC6        & $-1.40_{-0.04}^{+0.07}$ & $-1.7 \pm 0.3$ & $ ...           $ & ...               & ...                       \\
SC7        & $-0.61_{-0.04}^{+0.03}$ & $-1.0 \pm 0.4$ & $ ...           $ & ...               & ...                       \\
\hline
\end{tabular}
\end{table*}

The reason for the discrepancy between the results presented in this
contribution and those by \citet{Hwang14} is unclear. It is known that
NGC~6822 contains young and more metal rich populations of stars, and
they could be the cause of the observed results if they were
contaminating the photometric apertures. However, repeating the
photometry with smaller apertures in order to exclude possible
contaminants did not change the results in Figure~\ref{fig:SSPN6822}.
In addition, the ages are recovered to be old which argues against
younger stellar populations contaminating the derived total
magnitudes. As already noted, the optical (V-I) colours presented by
\citet{hwang11} are always redder than those in this work. It is clear from
examining Figure~\ref{fig:SSPN6822} that neither our or their optical colours
are consistent with their low spectroscopic metallicities, at least when
interpreted using the PARSEC1.1 models. A colour-magnitude diagram analysis of
the resolved stellar populations in these GCs would certainly help to
resolve this issue, and might even indicate the reason for the mismatch.

Another effect to be considered is the internal reddening within
NGC~6822 itself. As discussed earlier, this galaxy contains a
significant amount of gas and dust, which, if not properly accounted
for could cause the clusters to appear redder and thus more metal-rich
than they actually are. Looking at Figure~\ref{fig:spatialN6822} this
seems to be unlikely since many GCs are located outside the extent of
the HI disc, including several which are inferred to have high
metallicities.

\section{Kinematics}
\label{s:kinematics}

\subsection{Radial velocity measurements}

We derived heliocentric radial velocities from the spectra obtained
with the 4-m class telescopes via a customized $\chi^2$ minimisation
routine, using a number of appropriate radial velocity standard stars
and bright M31 GCs as templates \citep[see][]{Veljanoski14}. The derived
velocity from each science spectrum is the average of all minimisations
between that spectrum and the applied velocity templates. A full
description of the minimisation process, as well as the radial velocity
standard stars and clusters is presented in \citet{Veljanoski14}.
The $\chi^2$ minimisation method is similar to the more frequently used
cross-correlation technique \citep{TonryDavis79}, with the advantage
that it uses the uncertainties in the spectra of both the target and
template objects, which helps to weed out spurious peaks in the
$\chi^2$ functions and reduces the resulting velocity uncertainties.

During the course of the Keck observations no radial velocity standard
stars were observed.  However, the bright cluster SC7 was observed
with both the \emph{KPNO} and the \emph{WHT} telescopes, essentially
providing three independent velocity estimates that all agree with
each other very well (one of these comes from \emph{KPNO}, while two
come from the red and blue arms of the \emph{ISIS} spectrograph on the
\emph{WHT}). Thus we use our Keck observation of this cluster as a
radial velocity template against which the other Keck spectra were
matched, and adopt as its velocity the error weighted mean of the
three independent measurements obtained from the 4-m data sets.  The
same $\chi^2$ algorithm was used to estimate the radial velocity of
the Keck targets.

From all these data we have up to three independent velocity estimates
for 6 of the 8 GCs in NGC~6822, two of which (SC6 and SC7) possess no
previous velocity information.  Table~\ref{tab:rvN6822} lists the
velocity estimates coming from the different datasets and their
associated uncertainties.  Since these all agree within the
measurement uncertainties, we combine them into a single value by
performing an error weighted average and use these values for the
forthcoming analysis. For comparison, Table~\ref{tab:rvN6822} also
lists the velocity information available in the literature, as well as
the systemic velocity of NGC~6822. Our measurements are generally in
good agreement with those from the literature, while possessing
considerably smaller uncertainties.

Of the remaining two clusters, SC5 is sufficiently faint that to date
there have been no successful velocity measurements made for this
object. \citet{Hwang14} present a velocity measurement for SC4 (see
Table~\ref{tab:rvN6822}); however the uncertainties on their estimate
are large.  Although we did not obtain a longslit spectrum of SC4
during any of our observing runs, individual stars within this object
were targeted with \emph{DEIMOS} in multislit mode during our Keck
run. The purpose of these observations was to investigate the internal
kinematics of this rather extended cluster, and this analysis will be
presented elsewhere (Mackey et al., 2015, in prep.). Here we adopt the
outcome of a computation to determine the most-likely radial velocity
of this cluster given individual measurements for seven member stars,
$v = -75 \pm 3$\ km$\;$s$^{-1}$.

\begin{table*}
\centering
\scriptsize
\caption{Heliocentric radial velocities for the GCs around NGC~6822 as observed from each
  instrument. Also shown are the combined radial velocities which are then used for the kinematic
  analysis. Note that SC7 was used as a velocity template for the Keck data set.
  For reference, the systemic velocity of NGC~6822, as well as velocity information
  available in the literature are also shown.}
\label{tab:rvN6822}
\begin{tabular}{llllllll}
\hline
\hline
ID             & $KPNO\ v$         & $WHT_{blue}\  v$  & $WHT_{red}\ v$    & $KECK\ v$         & combined $v$      & Literature $v$         & Reference for                 \\
               & [km$\;$s$^{-1}$]  & [km$\;$s$^{-1}$]  & [km$\;$s$^{-1}$]  & [km$\;$s$^{-1}$]  & [km$\;$s$^{-1}$]  & [km$\;$s$^{-1}$]       & literature data               \\
\hline
NGC~6822       &  ...              & $ ...      $      & $ ...      $      & $ ...      $      & $ ...      $      & $ -57 $                & \citet{McConnachie12}         \\
Hubble VII     & $ -68\pm12 $      & $ ...      $      & $ ...      $      & $ ...      $      & $ -68\pm12 $      & $ -52; -65\pm20 $      & \citet{Cohen98,Hwang14}       \\
SC1            & $ -69\pm16 $      & $ ...      $      & $ ...      $      & $ -67\pm4  $      & $ -67\pm4  $      & $ -61\pm20 $           & \citet{Hwang14}               \\
SC2            & $...       $      & $ -72\pm12 $      & $ -80\pm7  $      & $ -74\pm5  $      & $ -76\pm4  $      & $ -106\pm31 $          & \citet{Hwang14}               \\
SC3            & $ -83\pm14 $      & $ ...      $      & $ ...      $      & $ ...      $      & $ -83\pm14 $      & $ -71\pm17 $           & \citet{Hwang14}               \\
SC4            & $ ...  $          & $ ...      $      & $ ...      $      & $ ...      $      & $ ...      $      & $ -115\pm58; -75\pm3 $ & \citet{Hwang14}; Mackey et al. (in prep.) \\
SC5            & $ ...  $          & $ ...      $      & $ ...      $      & $ ...      $      & $ ...      $      & $ ...  $               &                               \\
SC6            & $ -16\pm13 $      & $ ...      $      & $ ...      $      & $ -5\pm3   $      & $ -6\pm3   $      & $ ...  $               &                               \\
SC7            & $ -42\pm17 $      & $ -36\pm11 $      & $ -37\pm2  $      & template          & $ -37\pm2  $      & $ ...  $               &                               \\
\hline
\end{tabular}
\end{table*}

\subsection{Bayesian kinematic modelling}

Constraining the kinematic properties of the GCs around NGC~6822 is
one of the main goals of this paper. Unlike in massive galaxies, the
GC systems of dwarfs are typically populated by only a few members, so
it is important to use a method which will extract the maximum
information from the sparse data that are available. Following the
example from \citet{Veljanoski14}, we work in the Bayesian framework,
which enables us to construct a kinematic model and deduce the most
likely values for the quantities of interest.

Given the known rotation of both the HI disk and the spheroid (as
traced with carbon stars), we particularly want to test whether the GC
system of NGC~6822 exhibits any measurable degree of rotation. We thus
construct a kinematic model ${\cal M}$ which features both rotational
and velocity dispersion components. The rotation component is
constructed following the prescription from \citet{Cote01}:

\begin{equation}
v_{\rm p}(\theta) = v_{\rm sys} + A\,{\rm sin}(\theta - \theta_{0})
\label{eq:rot}
\end{equation}
where $v_{\rm p}$ are the measured radial velocities, $\theta$ and
$\theta_0$ are the position angles of the GCs and the GC rotation axis
respectively and $A$ is the rotation amplitude. The systemic radial
motion of the GC system taken to be the heliocentric motion of
NGC~6822, is labelled as $v_{\rm sys}$. As elaborated in
\citet{Cote01}, this simple motion along circular orbits assumes that
the GC system being studied is spherically distributed, that the
rotation axis lies in the plane of the sky and that the projected
angular velocity is a function of radius only.

Looking carefully at Figure~\ref{fig:spatialN6822} and
Table~\ref{tab:rvN6822}, it is noticeable that even though the mean
velocity of the GCs, which is $-59$~\kms, is similar to the systemic
motion of NGC~6822, there is a preference for the GCs to have radial
velocities more negative than the systemic velocity of their host.
Thus we retain $v_{\rm sys}$ as a free parameter instead of fixing in
to the known systemic velocity of NGC~6822.

The velocity dispersion is assumed to possess a Gaussian form, and not
to change as a function of projected radius. It is mathematically
described as:
\begin{equation}
\sigma^2  = (\Delta v)^2 + \sigma_0^2
\label{eq:constdisp}
\end{equation}
where $\sigma_0$ is the intrinsic velocity dispersion of the NGC~6822
GC system while $\Delta v$ are the uncertainties in the radial
velocity measurements.

Combining these two components, our kinematic model $\cal M$ takes the
following form:
\begin{equation}
p_{i,{\cal M}}(v_i, \Delta v_i|v_{\rm p}, \sigma) = \frac{1}{\sqrt{2\pi\sigma^2}}  \:\: e{^{- \frac{(v_i - v_{\rm p})^2}{2\sigma^2}}}
\label{eq:rotmodel-dE}
\end{equation}
where, as before, $v_{p}$ is the systemic rotation described by
Equation~\ref{eq:rot}, $v_i$ are the observed heliocentric radial
velocities of the GCs as presented in Table~\ref{tab:rvN6822}, and
$\sigma$ is a velocity dispersion as modelled by
Equation~\ref{eq:constdisp}. Hence, the likelihood function for the
above kinematic model is:
\begin{equation}
p_{{\cal M}}(D|\Theta) = {\cal L}_{{\cal M}}\left(v, \Delta v, \theta | A,\theta_0,v_{\rm sys},\sigma_0 \right)  = \prod_i p_{i,{\cal M}} 
\label{eq:like-dwarfs}
\end{equation}
where $v,\, \Delta v,$ and $ \theta$ are the observed properties of the GCs, while
$A,\, \theta_0, v_{\rm sys}$ and $\sigma_0$ are the model free parameters. We assume all priors for
the free parameters to be flat.

Even though our model features four free parameters, we systematically
calculate the likelihood function via a ``brute-force'' method over a
regular grid shaped by the parameter space of the kinematic model. The
rotation amplitude $A$ and the velocity dispersion parameters
$\sigma_0$ range from 0 to 180~\kms\ and from 0 to 120~\kms
respectively, both with a resolution of 2~\kms.  The axis of rotation,
$\theta_0$, is searched from 0 to $2\pi$~rad with a step size of
0.15~rad, while the $v_{\rm sys}$ ranges from -200 to 45~\kms\, with a
step size of 3~\kms.

With this set-up, however, it quickly became apparent that the
algorithm could not properly converge with the position angle
$\theta_0$ allowed to vary as a free parameter.  This is due to the
unusual linear configuration of the GCs and the small size of the
sample. Thus, in order to investigate the general kinematic properties
of this GC system, we consider three constrained models.  In the first
scenario, we assume that if the GC system is rotating, the rotation
axis is perpendicular to the GC alignment in the plane of the sky.
This is a reasonable assumption given that rotating configurations
usually form flattened ellipsoids or disks. We refer to this model as
to the ``disk model''; in this scenario the rotation axis would lie
close to that inferred for the spheroid by \citet{Demers06kin}. In the
second case we fix $\theta_0$ so that the rotation axis lies parallel
to the apparent alignment of the GCs. This rotating ``cigar model''
represents the case where the cluster system might share the rotation
exhibited by the HI disk. Finally we consider a scenario in which the
GCs have kinematic properties separate from those of the spheroid and
the HI disk, and exhibit no net rotation signal (i.e., $v_{\rm p}$ in
Equation \ref{eq:rotmodel-dE} is set to zero and the problem is
reduced to one with only two free parameters). This outcome would
match that inferred by \citet{Hwang14} from their kinematic study of
SC1-SC4.

\begin{figure*}
\centering
\includegraphics[width=.85\textwidth]{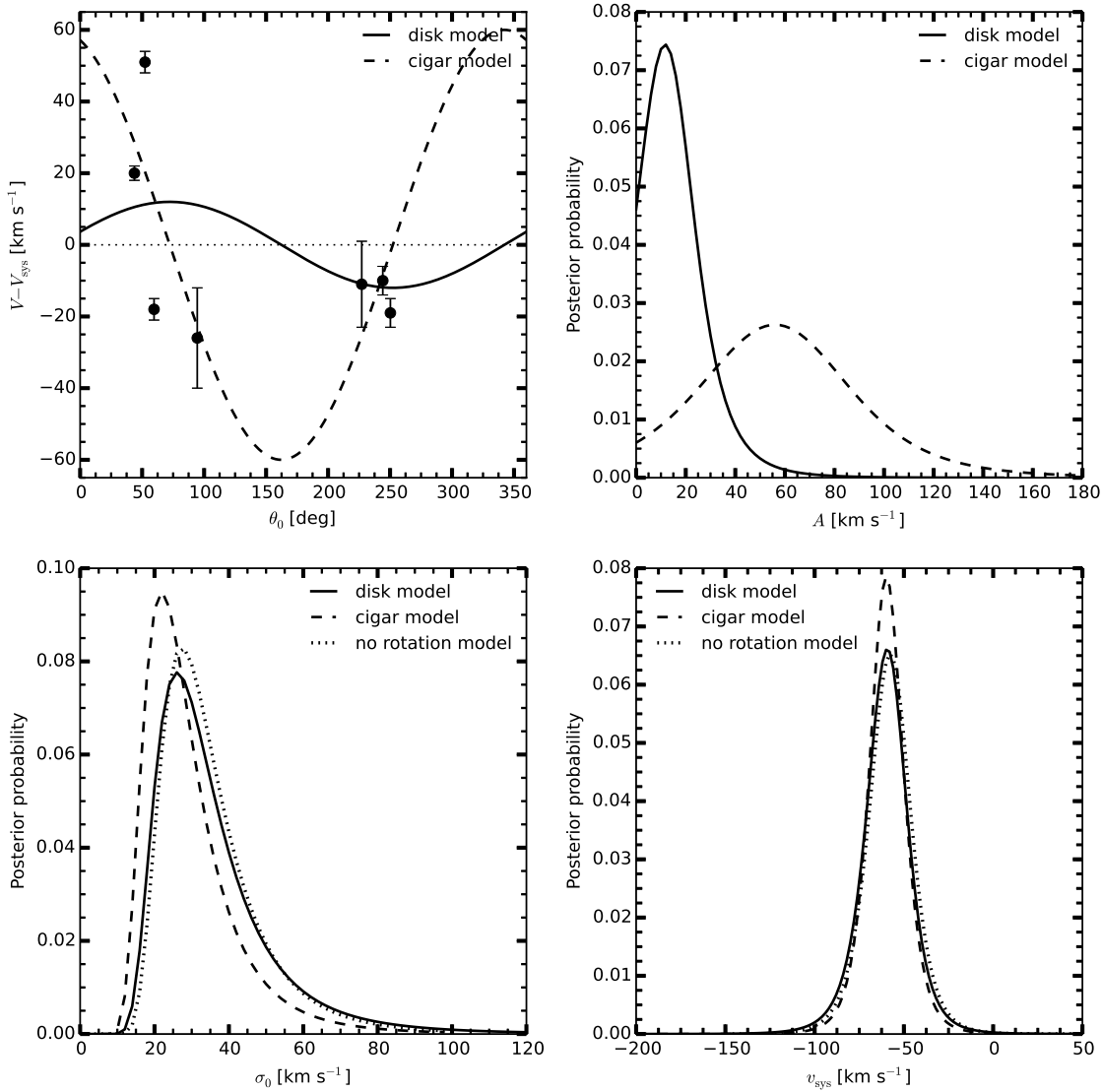}
\caption{Top left: heliocentric radial velocity, corrected for the
  systemic motion of NGC~6822, versus position angle of the GCs on the
  sky. The most-likely rotation curves from our two rotation-enabled
  models are displayed. The disk model shows evidence for a low degree
  of net rotation, while the cigar model detects a more considerable
  rotation signal (although this is likely to be non-physical -- see text).
  Top right: the marginalized posterior distribution functions for
  rotation amplitude. The amplitude for the disk model peaks at
  $12\pm10$~\kms, where the uncertainty corresponds to the 68\%
  confidence interval. The amplitude for the cigar model peaks at
  $56\pm31$~\kms. Lower left: the posterior functions for $\sigma_0$,
  with peaks between 22-27~\kms depending on the model used. In all
  three cases the distributions are quite similar.  Lower right: the
  posterior functions for $v_{\rm sys}$. The functions have peaks
  between -59 and -60~\kms, which is in excellent agreement with the
  systemic velocity of NGC~6822 found in the literature
  \citep{McConnachie12}}
\label{fig:kinN6822}
\end{figure*}

The results of our analysis are shown in Figure~\ref{fig:kinN6822}.
The general kinematic picture for this GC system is best described in
the top left panel, where the radial velocities of the GCs, corrected
for the systemic motion of NGC~6822, are plotted against their
position angles.  In the same panel the best-fit rotation curves are
overplotted, as described by Equation \ref{eq:rot}, from our two
scenarios which allow for rotation. The top right panel shows the
marginalized posterior probability distribution functions for the
amplitude in these two rotation-enabled models. From these two panels,
one can see that the disk model detects a low signature of net
rotation, where the most likely amplitude is $12\pm10$~\kms\ with 68\%
confidence. The probability of not having any net rotation signal is
moderately high according to this model.

For the cigar model, on the other hand, a strong coherent rotation
motion is inferred for the GC system.  The most likely amplitude is
$56\pm31$~\kms with 68\% confidence. This is quite a surprising
result, since any rapidly rotating system ought naturally to form a
flattened ellipsoid or disk structure. One possibility is that we are
observing a dynamically young system; however, there is at present no
evidence that this is true for the NGC~6822 GCs -- for example one
might expect to see a stellar stream or other tidal debris associated
with the clusters. A small amount of experimentation revealed that
this rotation signal is most likely a false detection driven by SC3 on
the one hand, and SC6 and SC7 on the other.  These objects have strong
deviations from the mean systemic velocity in opposite directions, and
happen to lie either side of the rotation axis when it is arranged
parallel to the GC alignment.  A small change in the assumed value of
$\theta_0$ in either direction strongly decreases the rotation signal.
Because there are only a handful of GCs in the entire system, these
few points are able to substantially influence the outcome of the
calculation (and indeed we believe it was this numerical instability
as $\theta_0$ moves across the GC alignment that led our original
four-parameter analysis to fail).

The lower left and right panels of Figure~\ref{fig:kinN6822} show the
marginalized posterior distributions of the velocity dispersion
$\theta_0$ and the systemic velocity of NGC~6822 $v_{\rm sys}$.
Regardless of the assumed model, the most likely value of $v_{\rm
  sys}$ is always found to be between -59 and -60~\kms with 68\%
confidence intervals of up to $\pm13$~\kms. This is in excellent
agreement with the literature value for the systemic velocity of
NGC~6822 from \citet{McConnachie12}. The velocity dispersion is also
well defined and does not show a large variation depending on the
assumed model. Its most likely value is $22_{-16}^{+10}$,
$26_{-18}^{+12}$, $27_{-19}^{+11}$~\kms for the cigar model, disk model,
and no-rotation model, respectively.

\section{The Dynamical Mass of NGC~6822}

GCs have a long history of being used as probes for estimating the dynamical
masses of the galaxies they surround \citep[e.g.][]{Federici90,Federici93,
Evans00}. We estimate the mass of NGC~6822 using its GCs as dynamical tracers
following a common approach \citep[e.g.][]{Woodley10,Veljanoski13a,
Veljanoski14}. Since we considered three kinematic scenarios, we will use the
results of each to estimate the mass of this dwarf galaxy. Solving the Jeans
equations \citep{BT87} is one common approach. Since our rotation-enabled models
found a varying degree of rotation in NGC~6822's GC system, the total mass of
this dwarf galaxy can be thought of as the sum from two components: a rotating
component, and a non-rotating pressure-supported component.

In the scenarios where a net rotation signature is detected, the
rotation component, $M_{\rm r}$, is simply calculated via:

\begin{equation}
M_{\rm r} = \frac{R_{\rm max}A^2}{G}
\end{equation}
where $A$ is the rotation amplitude of the GC system, $R_{\rm max}$ is
the projected radius of the outermost constituent, and $G$ is the
gravitational constant.

The pressure supported mass component is determined using the `Tracer
Mass Estimator' (TME) devised by \citet{Evans03}. The TME is
mathematically described as:

\begin{equation}
M = \frac{C}{GN}\sum\limits_{i=1}^{N}\left(v_i-v_{\rm sys}\right)^{2}R_i
\label{eq:TME}
\end{equation}
where $R$ is the projected radius from the centre of NGC~6822 for a
given GC, $v$ is the radial velocity of that GC, corrected for global
rotation of the system, while $v_{\rm sys}$ is the systemic radial
velocity of NGC~6822 itself. The index $i$ loops over each GC in the
sample of $N$ GCs that have available radial velocities. The constant
$C$ is dependent on the shape of the underlying gravitational
potential, $\alpha$, assumed to be scale-free; the radial distribution
of the mass tracers, assumed to be a power law with index $\gamma$;
and the anisotropy of the system.  For an anisotropic system, it takes
the form:

\begin{equation}
C = {4(\alpha\!+\!\gamma) \over \pi}
   {4\!-\!\alpha\!-\!\gamma\over 3\!-\!\gamma}
   {1\!-\!(r_{\rm in}/r_{\rm out})^{3-\gamma} \over 1\!-\!(r_{\rm in}/r_{\rm out})^{4\!-\!\alpha\!-\!\gamma}}.
\label{eq:C}
\end{equation}

For $\gamma = 3$, which is typical for a spheroidal stellar halo, $C$ becomes:

\begin{equation}
C = {\ds 4(\alpha\!+\!3)(1\!-\!\alpha) \over \ds \pi}
{\ds \log (\rout/\rin)\over \ds 1\!-\!(\rin/\rout)^{1\!-\!\alpha}}.
\label{eq:Cg3}
\end{equation}

We consider cases where the slope of the underlying gravitational
field, $\alpha$, is set to 0, which assumes an isothermal halo
potential for NGC~6822, and to 0.55, as for a NFW profile
\citep{NFW96,Watkins10}. For $\rin$ and $\rout$ we adopt the smallest
and largest projected radii, respectively, exhibited by the GCs in our
sample. As before, we include the velocity measured for SC4 by Mackey
et al. (in prep.); however, we now exclude Hubble-VII from the
analysis. If a tracer object having a projected radius of zero is used
in the TME, it will produce a singularity in Equation~\ref{eq:Cg3}.
Hubble-VII has a projected radius of just 0.13~kpc, which causes the
constant $C$ to increase anomalously.

\begin{table*}
\centering
\scriptsize
\caption{Dynamical mass estimates for NGC~6822. See text for details.}
\label{tab:N6822-new}
\begin{tabular}{@{}lcccccc@{}}
\hline
\hline
Model       & N$_{GC}$ & $\alpha$ & M$_p$ [\Msun]            & M$_r$ [\Msun]           & M$_{total}$ [\Msun]      & M/L[\Msun/L$_{\odot}$] \\
\hline
Disk        & 6        & 0        & $4 \pm 2 \times 10^{9}$  & $3 \pm 6 \times 10^{8}$ & $4 \pm 2 \times 10^{9}$  & 40 \\
Disk        & 6        & 0.55     & $3 \pm 1 \times 10^{9}$  & $3 \pm 6 \times 10^{8}$ & $4 \pm 1 \times 10^{9}$  & 40 \\
cigar       & 6        & 0        & $4 \pm 2 \times 10^{10}$ & $7 \pm 8 \times 10^{9}$ & $4 \pm 2 \times 10^{10}$ & 400 \\
cigar       & 6        & 0.55     & $3 \pm 1 \times 10^{10}$ & $7 \pm 8 \times 10^{9}$ & $4 \pm 2 \times 10^{10}$ & 400 \\
No rotation & 6        & 0        & $4 \pm 1 \times 10^{9}$  &                         & $4 \pm 1 \times 10^{9}$  & 40 \\
No rotation & 6        & 0.55     & $3 \pm 1 \times 10^{9}$  &                         & $3 \pm 1 \times 10^{9}$  & 30 \\
\hline
\end{tabular}
\end{table*}

Table~\ref{tab:N6822-new} lists estimated masses for NGC~6822 within
$\sim 11$~kpc, i.e. the projected radius of the most remote cluster
(SC1). The uncertainties of the pressure-supported mass component are
calculated via the jackknifing technique, while the uncertainties in
the rotation component are only due to the error propagation of the
detected amplitude in the relevant scenarios. Total massess estimated
assuming the rotating disk and the non-rotating scenario yield similar
results, which is unsurprising given the low degree of rotation
detected in the disk model. For these cases our measurements yield an
estimated mass of $3-4 \times 10^{9}~M_{\odot}$, with a corresponding
mass-to-light ratio of $30-40$. These estimates are, however, smaller
than those from \citet{Hwang14} who found the mass of NGC~6822 within
11~kpc to be $7.5_{-0.1}^{+4.5} \times 10^9$~\Msun using four extended
clusters only (SC1-SC4). The discrepancy in the estimates is likely
due to the large uncertainties in the velocity measurements, in
combination with their smaller cluster sample. Still, our inferred
mass-to-light ratio makes NGC~6822 a highly dark matter dominated system
when compared to other dwarf galaxies of similar luminosity
\citep[e.g.][]{Mateo98,McConnachie12,Kirby14}.

On the other hand, assuming the rotating cigar scenario we estimate the mass of
NGC~6822 to be unreasonably high, about $4 \times 10^{10}~M_{\odot}$, as
indicated by the associated mass-to-light ratio. This adds further confidence
that the high rotation signal found for this model is indeed a spurious
detection.

It is worth acknowledging the caveats regarding the choice of $\alpha$
and $\gamma$ parameters which feature in the TME. It is unclear
whether their assumed values are indeed the right choices when
estimating the mass of NGC~6822, thus adding additional uncertainty
which is not formally included. Using the same method and identical
assumptions, but only 4 extended GCs, \citet{Hwang14} found the total mass
of NGC~6822 enclosed within 11 kpc to be
$7.5^{+4.5}_{-0.1} \times 10^9$~\Msun. This most likely indicates that
when dealing with a small samples such as this one, the inclusion or
exclusion of a single data point can have an important effect on the
results. For reference, using the kinematics of the HI gas,
\citet{Weldrake03} found a mass of $3.2 \times 10^9$~\Msun\, within
$\sim 5$~kpc from the centre of NGC~6822.

\section{Discussion}

GCs hosted by dwarf galaxies have been often found to have similar colours as
GCs residing in the haloes of massive galaxies \citep[e.g.][]{KunduWhitmore01a,
Larsen01,Peng06}, and our results are in keeping with that. Comparing the
optical colours of the GCs hosted by NGC~6822 with those located in the M31
outer halo, it is found that they are all mutually consistent,
having $(V-I)_0 \sim 0.9$. This is also true for the GCs residing in the
dwarf elliptical galaxies NGC~147 and NGC~185, located in the outer halo of
M31 \citep{Veljanoski13b}. In terms of their number, extended GCs dominate the
population of NGC~6822 and there is a similarity between the mean absolute
magnitudes of the extended GCs hosted by NGC~6822 and by M31, which are found
to be $M_{V_0} = -5.9 \pm 0.7$ and $M_{V_0} = -5.6 \pm 0.7$, respectively. This
may either be a fundamental property of extended GCs in general, or is a result
of a bias arising from the faint limit of the survey imaging from which these
objects were discovered. Given that some of them lie right at the detection
limit, it is possible that the extended GCs have a larger range of absolute
magnitudes than observed at present, but the currently available data allows
only for the bright end of the luminosity function to be observed, resulting in
the low spread and similar $M_{V_0}$ values. Nonetheless, the range of
luminosities, structures, ages and metallicities spanned by the GCs in NGC6822
is broadly consistent with that observed in the M31 halo population
\citep[e.g.][]{Mackey06,Mackey07,Mackey10a,Mackey13,Mackey13b,Alves-Brito09,
Sakari15}.

A number of studies have focused on GC systems hosted by dIrr galaxies
outside the Local Group
\citep[e.g.][]{Seth04,Sharina05,Georgiev06,Georgiev08}. They have
found that GCs have typical optical colours within the range of $0.8 <
(V-I)_0 < 1.1$, consistent with our measurements.  In their study,
\citet{Georgiev08} searched for GCs in 19 nearby (2-8 Mpc) dIrr
galaxies using archival \emph{HST} data. The galaxies in their sample
are members of dwarf galaxy associations only, without a dominant
massive galaxy nearby. In the final sample of GC candidates that
passed all their selection criteria, \citet{Georgiev08} found an
absence of objects having $(V-I)_0 < 1$ and $M_{V_0} \gtrsim -6$, i.e.
faint and blue GCs. The authors claim that this effect is not due to a
bias in their GC selection criteria, nor a bias due to the depth of
the imaging, since they are able to identify clusters down to $M_{V_0}
\simeq -4$. In NGC~6822 however, there are 2 extended GCs which
satisfy these criteria. If such clusters were present in the sample of
\citet{Georgiev08}, they were most likely missed because of their
low surface brightness as well as their extended morphology.

We have considered three scenarios for the kinematic configuration of
the GC system of NGC~6822.  As indicated in the previous Section, the
so-called ``disk'' and ``no-rotation'' models are the most plausible
of these.  In the disk scenario our analysis provides tentative
evidence for a weak rotation signal of amplitude $12\pm 10$ \kms,
although clearly the case of zero rotation is not strongly excluded.
It is interesting to note that this putative rotation would be in the
same sense as that detected for stars in the underlying stellar
spheroid by \citet{Demers06kin}, and also of a comparable magnitude
(although we note that the Demers et al. measurements do not extend to
particularly large galactocentric radii). Similarly, the velocity
dispersion we infer for the GC system matches closely that determined
from the \citet{Demers06kin} kinematic sample, once the global
rotation has been accounted for \citep[see][]{Hwang14}. These
similarities suggest that the GC system of NGC~6822 is plausibly
associated with its stellar spheroid, and indeed this would be
consistent with the clear linear alignment of the GCs with the major
axis of the spheroid.

Our third kinematic model -- the so-called ``cigar'' model -- would
imply a large systemic rotation amongst the GC population, although
this is strongly sensitive to the precise orientation of the rotation
axis. In this case, the sense of rotation would be opposite to that
seen for the HI gas by \citet{deBlok00}; furthermore, the inferred
amplitude would be considerably larger than that seen for the HI gas
at comparable distances along its major axis to those of the GCs.  In
the event this model was correct, it would appear that the GCs are not
asscociated with the HI gas component of NGC~6822.  The GCs could
constitute a dynamically young system, in which case a search for a
stellar stream along the major axis of the NGC~6822 spheroid may prove
fruitful.

\section{Summary}

In this paper we present a uniform optical and near-IR photometric
study of the GC system hosted by NGC~6822. The photometric
measurements are used to estimate the ages and metallicities of the
GCs via a colour-colour plot and an empirical colour-metallicity
relation. All GCs are found to have ages consistent with 9~Gyrs or
older, in agreement with past studies. On the other hand, the clusters
are found to exhibit a range of metallicities (${\rm -1.6 \lesssim
[Fe/H] \lesssim -0.4}$), at odds with previous spectroscopic studies
which found them to be very metal-poor ([Fe/H]$<-2.0$).  Metallicity
measurements from resolved colour-magnitude diagrams would alleviate
this tension, as would higher S/N spectroscopy.

We also modelled the kinematics of the GC system using new
velocities for six members, two of which had no previous spectroscopic
information. With the caveat at our results are based on a small
sample, we found tentative evidence for a weak rotation signal
of amplitude $12 \pm 10$~\kms\ in the case where the rotation axis
sits perpendicular to the linear arrangement of the GCs on the sky.
However, zero rotation is not strongly excluded by
our analysis. If the rotation is real, then it would be in the same
sense as inferred for the underlying stellar spheroid by
\citet{Demers06kin}, and of a comparable amplitude. Furthermore, the
velocity dispersion we determine for the GC system, $\sigma_0 =
26^{+18}_{-12}$~\kms is very similar to that exhibited by stars in the
spheroid \citep[see][]{Hwang14}. These similarities suggest to us
that the GCs are plausibly associated with the spheroidal component of
NGC~6822.

Finally, using the GCs as dynamical mass tracers, the total mass of
NGC~6822 is recalculated to be roughly $(3-4) \times 10^{9} M_{\odot}$,
subject to various systematic effects due the small data sample and the
implicit assumptions built into the TME. The corresponding mass-to-light
ratio sits in the range $\sim 30 - 40$, implying that NGC~6822 hosts a
substantial dark matter component.

\section*{Acknowledgments}
JV and AMNF acknowledge support from an STFC Consolidated Grant
awarded to the IfA. JV acknowledges partial support from NOVA.
ADM is grateful for support by an Australian Research Fellowship
(Discovery Projects Grant DP1093431) from the Australian Research
Council. APH was partially supported by Sonderforschungsbereich
SFB 881 ``The Milky Way System" of the German Research Foundation.
We are thankful to Alexander B. Rogers for parallelizing our
kinematics code. We are also thankful to Erwin de Blok for kindly
providing the data for the HI map used in Figure~\ref{fig:spatialN6822}.

The WHT is operated on the island of La Palma by the Isaac Newton
Group in the Spanish Observatorio del Roque de los Muchachos of the
Instituto de Astrof\'{i}sica de Canarias.

\bibliographystyle{mn2e}

\bsp

\label{lastpage}

\end{document}